\documentclass[twocolumn,showpacs,floatfix,epsfig]{revtex4}
\usepackage[final]{graphicx}
\usepackage{epsfig}

\begin{document}

\title{Bounds on the bound $\eta^3$He system}
\author{A. Sibirtsev$^1$, J. Haidenbauer$^1$,
J.A. Niskanen$^2$, Ulf-G. Mei{\ss}ner$^{1,3}$}

\affiliation{
$^1$Institut f\"ur Kernphysik (Theorie), Forschungszentrum J\"ulich,
D-52425 J\"ulich, Germany \\
$^2$Department of Physical Sciences, PO Box 64, FIN-00014 University of
Helsinki, Finland \\
$^3$ Helmholtz-Institut f\"ur Strahlen- und Kernphysik (Theorie),
Universit\"at Bonn \\
Nu\ss allee 14-16, D-53115 Bonn, Germany}

%\date{\today}

\begin{abstract}
We investigate the relation between the $\eta^3$He
binding energy and width and the (complex) $\eta^3$He scattering
length. Following our systematic analysis of the $\eta^3$He
scattering length we set limits on the $\eta^3$He binding.
If bound states exist the binding energy (width) should not exceed
5 MeV (10 MeV). In addition, we comment on a recently
claimed observation of an $\eta$ mesic $^3$He quasi-bound state 
by the TAPS collaboration based on $\eta$ photoproduction data. 
Although our limits are in reasonable agreement with the values 
reported by this collaboration, our analysis of these data does not 
lead to a solid conclusion concerning the existence of an eta-mesic
bound state. More dedicated experiments are necessary for further 
clarification.

\vspace{-7.5cm}\hfill{\tiny  FZJ-IKP-TH-2004-11, HISKP-TH-04/15}\vspace{7.3cm}

\end{abstract}
%\\ Keywords:  one; two; three}

\maketitle

The formation of $\eta$-nucleus quasi-bound states has been
investigated for a long time. While no such states have
been directly observed, a quantity closely related to the
existence of bound states, namely the $\eta$-nucleus scattering
length has been intensively studied both experimentally and
theoretically. By factorizing the strongly energy dependent
final state interaction part from the slowly energy dependent production
mechanism~\cite{Watson,Migdal}, one may hope to extract the
scattering length from production reactions~\cite{Mayer}.
However, that information is necessarily indirect, since the
quantity actually related to binding, namely the sign of the real
part $\Re a_{\eta {\rm He}}$ of the complex scattering length,
cannot be determined from the cross section of the reactions 
$pd{\to}\eta^3{\rm He}$ and $\pi^+t{\to}\eta^3{\rm He}$~\cite{Green,Sibirtsev}.
One may note that there is a presumably quite difficult experimental
possibility of using charge symmetry breaking $\eta\pi^0$ meson-mixing
in pion production in the neighborhood of the $\eta$
threshold to do this~\cite{Baru}.
On the other hand, several theoretical studies attempt to calculate
$\eta$-nucleus scattering starting from the elementary $\eta N$
interaction using for example optical models based on the impulse
approximation~\cite{Wilkin}, sophisticated versions of
multiple scattering expansions~\cite{Wycech,Rakityansky} or even
Faddeev-Yakubovsky equations~\cite{Fix1}. Although 
these calculations are quite contradictory in numerical details,
some of them still indicate a possibility of a large negative
$\Re a_{\eta {\rm He}}$ -- the latter being considered as a smoking gun 
evidence for the existence of a $\eta ^3{\rm He}$ bound state. 
In contrast, the optical model
calculations of Refs.~\cite{Haider1,Liu,Chiang,Haider2} exclude
the possibility of the $\eta$  binding  in the three-nucleon system
and proposed~\cite{Haider2} that the lightest bound system would
be with $^4$He. Also a Skyrme model calculation suggests $^4$He
to give weak binding, while the case of $^3$He is inconclusive
\cite{Scoccola}
Recently our knowledge of the $\eta^3{\rm He}$ scattering length was 
revisited~\cite{Sibirtsev}
via a systematic analysis of presently available data on the
$pd{\to}\eta^3{\rm He}$ reaction. It is natural to extend this
study and to investigate in how far the limits deduced for the scattering 
length in Ref. \cite{Sibirtsev} provide also constraints on the 
binding energy for the $\eta^3{\rm He}$ system. In particular, a
phenomenological understanding of the relation between the scattering
length and the depth of the binding and the width of the state
would be valuable in planning possible experiments aimed for the
direct observation of bound states. In this paper we present a
calculation where both the depths and widths of the 
bound states together with the corresponding scattering length
are interconnected in terms of complex potentials.

The $\eta$-nucleus optical potential is taken to be proportional to
the density of the $^3$He nucleus,
\begin{equation}
V= - \frac{4\pi}{2\mu_{\eta{\rm He}}} (V_R + i V_I) \rho(r),
\end{equation}
for which the Gaussian form
\begin{equation}
\rho(r) = \frac{1}{(\sqrt\pi \alpha)^3}\, e^{-r^2/\alpha^2}\, ,
\quad \alpha = \sqrt{\frac{2}{3} \, \langle r^2 \rangle}
\end{equation}
corresponding to an root-mean-square radius 1.9~fm is adopted,
and $\mu_{\eta{\rm He}}$ is the reduced mass. 
Within the standard
optical model calculations the strength parameters $V_R$ and $V_I$
were taken as thrice the elementary $\eta N$ scattering length
adjusted with the ratio of the reduced masses of the $\eta$-nucleus
system and the $\eta N$ system as
\begin{equation}
 V_R + i V_I = 3 a_{\eta N}\,
\frac{\mu_{\eta {\rm He}}}{\mu_{\eta N}}\, .
\end{equation}
More precisely, this factor of three stems from the impulse approximation
underlying such an appraoch.
Here the sign definition of the scattering length $a$ is given by the
standard effective range convention in meson physics
\begin{equation}
q\cot\delta = \frac 1 a + \frac 1 2 r_0 q^2 + {\cal O} (q^4) \, , 
\end{equation}
where $\delta$ is the phase shift and $r_0$ is the effective range.
We should emphasize, however, that 
the present study is not an optical model calculation
in the above narrow sense.
Instead, here the strength parameters $V_R$ and $V_I$
are freely varied to study for which scattering lengths one
might expect bound states to exist and with which energies.
As a numerical check, the values of $V_R{=}2.235$~fm and
$V_I{=}1.219$~fm yield the $\eta^3{\rm He}$ scattering length
$a_{\eta{\rm He}}{=}-2.31{+}i 2.57$~fm in agreement with the
result of Ref.~\cite{Wilkin}.
\begin{figure}[t]
\psfig{figure=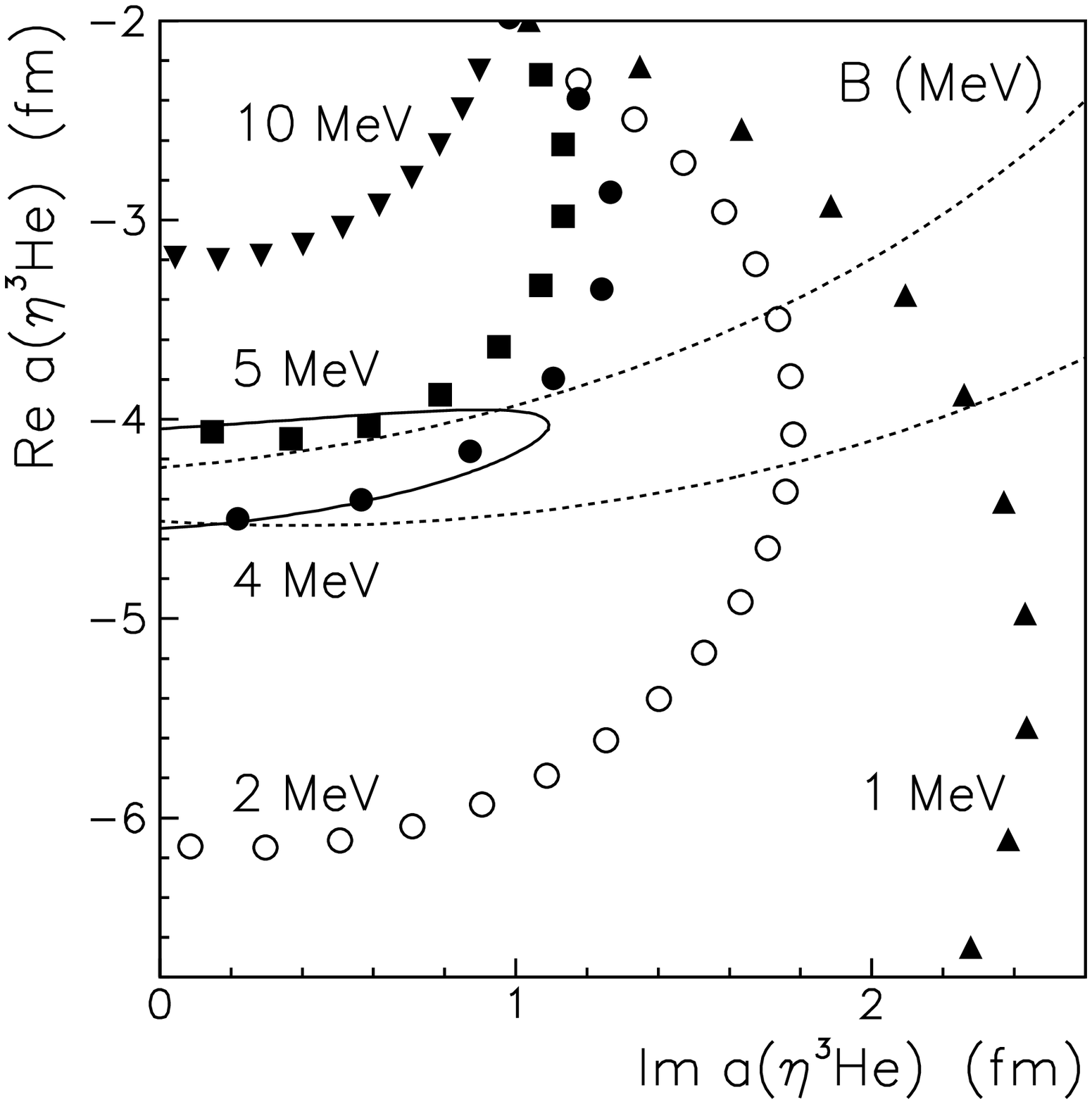,height=6.5cm.,width=8.4cm}\vspace*{-6mm}
\psfig{figure=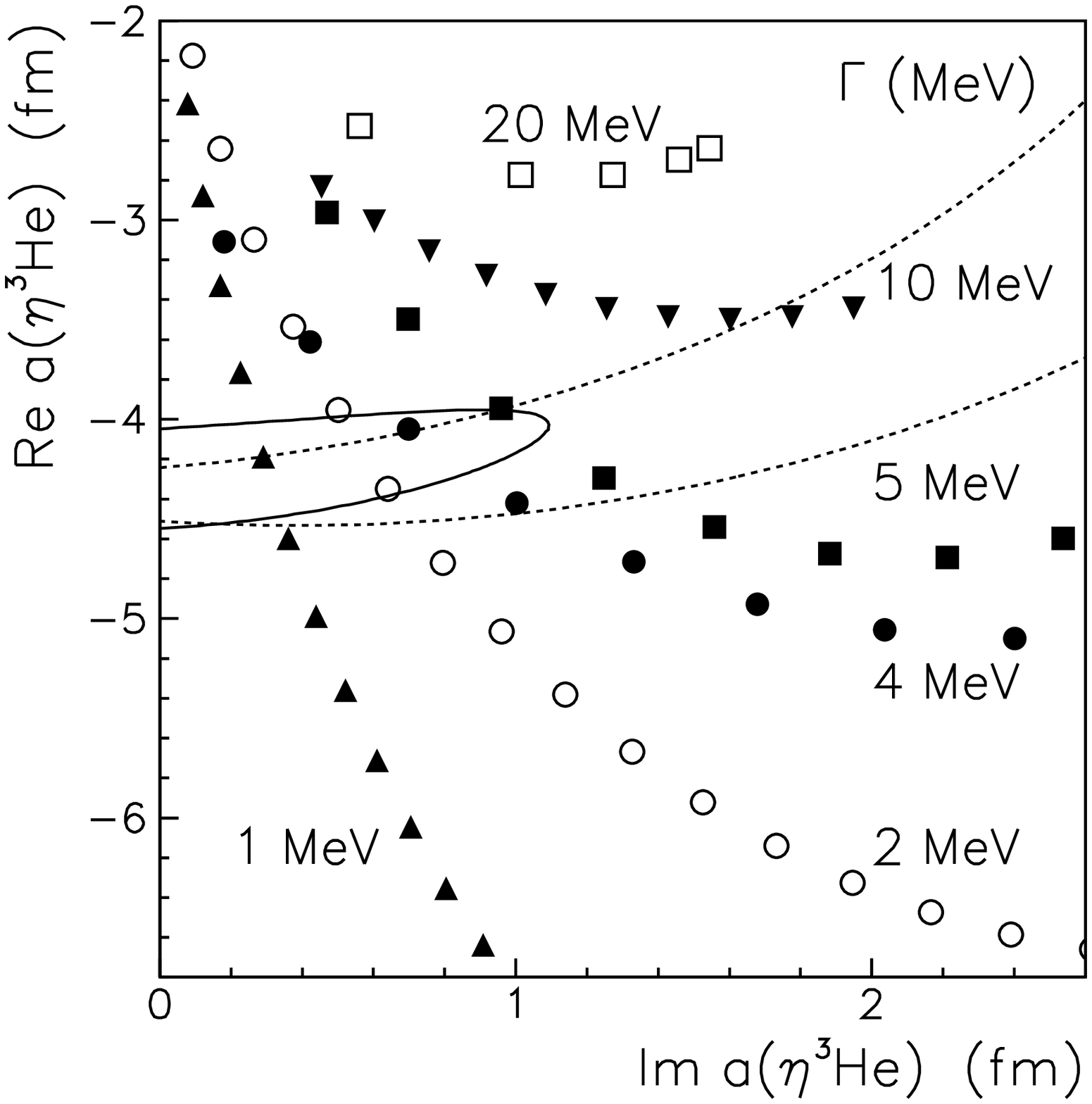,height=6.5cm.,width=8.4cm}
\vspace*{-4mm}
\caption{The $\eta^3{\rm He}$ binding energy $B$ (upper panel) and width
$\Gamma$ (lower panel) as functions of the imaginary and real parts
of the $\eta^3{\rm He}$ scattering length. Results are presented for 
$B$ and $\Gamma$ of 1 MeV (triangles), 2 MeV (open circles), 4 MeV
(close circles), 5 MeV (close squares), 10 MeV (inverse triangles)
and 20 MeV (open squares). The solid and dashed contours show our
$\chi^2{+1}$ solutions for the $\eta^3{\rm He}$ scattering
length~\cite{Sibirtsev} for the various data sets as explained in the text.}
\label{bin1a}
\end{figure}
For the scattering calculations the solution of the
Schr\"odinger equation with the asymptotic boundary condition
\begin{equation}
\psi_{0k}(r) \rightarrow \cos\delta\, j_0(kr) - \sin\delta\, n_0(kr)
\end{equation}
is standard and the numerics is accurate, i.e. better than 0.01~fm in the
relevant calculated quantities. The bound state is obtained from
the corresponding homogeneous  integral equation
\begin{equation}
\psi_E(r) = \int_0^\infty G_0(r,r') V_{\rm opt}(r') \psi_E(r')
r'^2 dr'
\end{equation}
by iterating to a self-consistent energy eigensolution. Here
the numerics is most difficult in the real and barely bound
region. Therefore, no results with less than 0.1~MeV binding
are actually used. In this worst situation still we would
consider the accuracy of our calculations to be better than 20~keV.
The above potential with varying real and imaginary
strengths is applied to calculate both the complex binding
energies and scattering lengths.
Fig.~\ref{bin1a} shows our results for the $\eta^3{\rm He}$
binding energy $B$  and width $\Gamma$ as functions of the imaginary and
real parts of the $\eta^3{\rm He}$ scattering length. The different
symbols indicate our results for $B$ and $\Gamma$
of 1, 2, 4, 5, 10 and 20 MeV. The solid and dashed contours
show our $\chi^2{+}1$ solutions for  the $\eta^3{\rm He}$ scattering
length~\cite{Sibirtsev}. The solid line is the solution obtained
by a simultaneous fit of the $pd{\to}\eta^3{\rm He}$ data from
Mayer et al.~\cite{Mayer} and Berger et al.~\cite{Berger}, while the
dashed line is our result evaluated from the data of
Mayer et al.~\cite{Mayer} alone.
The calculations shown in  Fig.~\ref{bin1a} indicate 
that, for example, the $V_R$ and $V_I$ strength of Ref.~\cite{Wilkin}
resulting in $a_{\eta{\rm He}}{=}{-}2.31+i2.57$~fm could not
provide binding of the $\eta^3{\rm He}$ system. Furthermore,
it is immediately clear that  the predictions for the $\eta^3{\rm He}$
scattering length from Refs.~\cite{Wilkin,Wycech,Rakityansky,Fix1}
do not lie in the region of binding, even though some of these studies
indicate support for bound states \cite{cracow} since they yield a
negative $\Re a_{\eta {\rm He}}$.
Looking at the results in Fig.~\ref{bin1a} in more detail
one  detects some strong dependencies. It seems
impossible to get binding under the condition
$\Im a_{\eta {\rm He}}{\ge}|\Re a_{\eta {\rm He}}|$. This finding is
in line with earlier expectations as conditions of quasi-bound states
for the scattering lengths formulated~\cite{Haider2}
as $-\Re a_{\eta {\rm He}}{>}\Im a_{\eta {\rm He}}{\ge}0$
or the more restrictive condition
$\Re [a_{\eta {\rm He}}^3(a_{\eta {\rm He}}^* - r_{0\eta {\rm He}}^*)] > 0$
involving also the effective range, given in Ref. \cite{Sibirtsev}.
Often these conditions are overlooked as criteria of bound states.
However, it should be also noted that neither of these conditions is
{\it sufficient} for the existence of a bound state. 
As an interesting feature it was found that
a very small imaginary strength $V_I$ could produce a large
imaginary part in the energy $E=-B-i\Gamma/2$ and scattering length
$a_{\eta {\rm He}}$ in the case of barely bound systems.
The effect is relatively
larger in the former by a factor of two. For example, for $V_R=2.1$ fm
giving only 0.25 MeV binding, a change of $V_I$ from zero to 0.1 fm
changes $B$ to 0.21 MeV and $\Gamma$ to 0.38 MeV, 
thus a 5\% imaginary strength produces essentially as large an
imaginary part in the energy as the real part. This may be
understood, if one considers that close to threshold
most of the (real) potential contribution is cancelled by the
kinetic energy. In the scattering length the above change is
``only'' $a_{\eta {\rm He}}=-17.1$ fm $\rightarrow$
$-14.3 +i6.0$ fm. However, close to the binding threshold also
changes in the scattering length are drastic, both in the real
and imaginary parts, a fact related to the loss of binding.
With deeper binding the imaginary part of the scattering length
becomes smaller. However, with an increasing imaginary strength
or  $\Im a_{\eta{\rm He}}$ the binding vanishes even
for strongly attractive potentials and
at the same time the state may become really very wide.
There is an accumulation point at about $-2.0{+}i0.9$~fm
to which all equal value contours converge with strengthening
potentials.
An even more important numerical finding was that, while
$E$ and $a_{\eta {\rm He}}$ are individually dependent on
the potential shape,
the relation between them was relatively model independent.
For two very different density profiles $\rho(r)$ the
differences between the complex energies were less than 10\%
for the {\it same} complex scattering lengths in the region
of interest. However, while
a well-known relation between the binding energy, scattering
length and effective range \cite{Joachain} holds quite well for
real binding energies of even 10 MeV, we found that it fails
for the complex case. Therefore, the numerically established only 
weakly model dependent relation cannot be expressed by a simple 
analytic formula.
Presently there are no data or any solid arguments to prove that
the real part of the $\eta^3$He scattering length actually is negative
and that the $\eta^3$He system should be bound.
But, to estimate the possible binding
energy $B$ and width $\Gamma$ we select our solution for
$a_{\eta {\rm He}}$ with a negative real part.
It may be noted that the result of our  analysis~\cite{Sibirtsev}
while taking  $\Re a_{\eta {\rm He}}{<}0$ would be within the binding
region as shown by the contour lines in Fig.\ref{bin1a}. In fact,
the standard rectangular error limits~\cite{Sibirtsev}
$a_{\eta^3{\rm He}}{=}({-}4.3{}{\pm}0.3){+}i(0.5{\pm}0.5)$ fm would suggest
a bound state with the binding energy $B{=}4.3{\pm}1.2$
and width  $\Gamma{=}2.8{\pm}2.8$ MeV. Taking our solution
shown by the dashed lines in Fig.\ref{bin1a} and obtained from
the data of Mayer et al.~\cite{Mayer} alone we deduce the limits for
$\eta^3$He binding energy $B{\le}$5~MeV and the width
$\Gamma{\le}$10~MeV.

\begin{figure}[t]
\psfig{figure=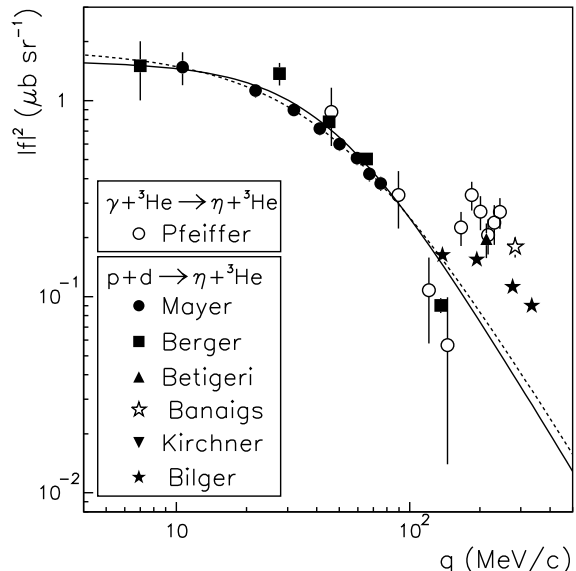,height=8.5cm.,width=8.5cm}\vspace*{-6mm}
\vspace*{-2mm}
\caption{Spin and angle averaged transition amplitudes
$|f|^2$ extracted from
$\gamma^3{\rm He}{\to}\eta^3{\rm He}$ \cite{Pfeiffer} and
$pd{\to}\eta^3{\rm He}$~\cite{Berger,Banaigs,Mayer,Bilger,Betigeri,Kirchner}
data  as functions of the final momentum $q$ in the c.m. system.
The $|f|^2$ from coherent photoproduction was arbitrarily multiplied
by a factor of 6. The
solid line is our overall fit~\cite{Sibirtsev} to low energy data from
by Mayer et al.~\cite{Mayer} and Berger et al.~\cite{Berger},
while the dashed line shows our fit to the data from
Mayer et al.~\cite{Mayer} alone.}
\label{bin3}
\end{figure}

\begin{figure}[tb]
\vspace*{-5mm}
\psfig{figure=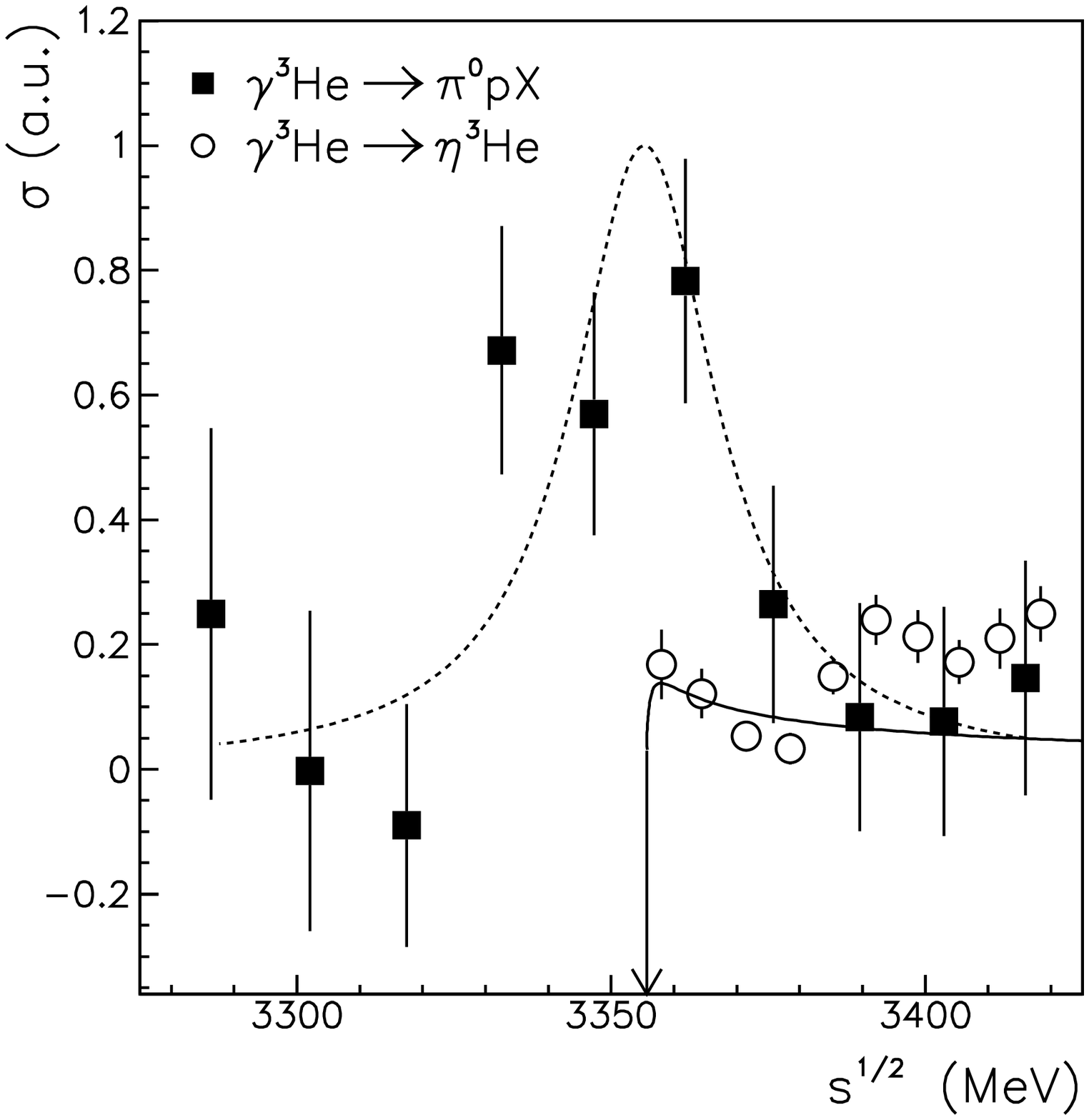,height=8.5cm.,width=8.5cm}\vspace*{-6mm}
\vspace*{-2mm}
\caption{The $\gamma^3{\rm He}{\to}\eta^3{\rm He}$ cross section
(open circles) and the  difference between the
$\gamma^3{\rm He}{\to}\pi^0pX$ reaction for
$170^0{\le}\theta_{\pi^0p}{\le}180^0$ and that for
$150^0{\le}\theta_{\pi^0p}{\le}170^0$ (solid squaress) as a function of the
invariant collision energy $\sqrt{s}$. The arrow indicates
the $\eta$ production threshold. The dashed curve is a fit to the data based 
on Eq. (\ref{Breit}) and the solid curve is the result based on 
Eq. (\ref{amp2}), employing the amplitude shown in Fig. \ref{bin3} (solid line).
}
\label{bin4}
\end{figure}

There are some new data on $\eta$ and $\pi^0$ photoproduction
on $^3$He from the TAPS Collaboration~\cite{Pfeiffer} at MAMI.
Fig.~\ref{bin3} shows the spin and
angle averaged squared transition amplitude $|f|^2$ 
extracted from data on 
$pd{\to}\eta^3{\rm He}$~\cite{Berger,Banaigs,Mayer,Bilger,Betigeri,Kirchner}
cross sections $\sigma$ \cite{Sibirtsev} as
\begin{equation}
|f|^2 =\frac{k}{q}\frac{\sigma}{4\pi},
\label{amp2}
\end{equation}
where $k$ and $q$ are the initial and final particle momenta in
center-of-mass (c.m.) system, together with corresponding results obtained from the
new TAPS data on $\gamma^3{\rm He}{\to}\eta^3{\rm He}$ \cite{Pfeiffer}.
The lines in Fig.~\ref{bin3} show the squared reaction
amplitude given by~\cite{Watson,Migdal}
\begin{equation}
|f|^2 =\left|\frac{f_p}{1-iaq}\right|^2,
\end{equation}
where $f_p$ is the $s$-wave production operator, assumed to
be independent of the final momentum $q$ near the reaction
threshold and $a$ is the complex $\eta^3$He scattering length.
The solid line in Fig.~\ref{bin3} shows our overall
fit~\cite{Sibirtsev} to low-energy data from
by Mayer et al.~\cite{Mayer} and Berger et al.~\cite{Berger},
while the dashed line shows our fit to the data from
Mayer et al.~\cite{Mayer} alone.
For $q{\le}100$~MeV, where the $\eta^3$He final state interaction
dominates, the coherent photoproduction data are in good agreement with
$pd{\to}\eta^3{\rm He}$ measurements and can be reasonably reproduced
by our $s$-wave  analysis~\cite{Sibirtsev}. In this sense the
$\gamma^3{\rm He}{\to}\eta^3{\rm He}$ data could not provide new
information about the sign of $\Re a_{\eta{\rm He}}$ and
from these data alone it is not possible to draw conclusions 
about the existence of the $\eta^3$He bound state.
However, Ref. \cite{Pfeiffer} also reports a small enhancement in
the $\gamma^3{\rm He}{\to}\pi^0pX$ reaction for the $\pi^0p$ $180^\circ$ 
opening angle spectra as compared with other opening angles 
for energies near the $\eta$ threshold. The enhancement 
is assumed to arise in particular from the pionic decay
of an $N^\ast (1535)$ resonance at rest in the $^3$He nucleus
so that its decay products should have opposite momenta in the
c.m. system. The $N^\ast (1535)$ resonance in turn is thought of being
formed by absorption of a bound $\eta$ meson on a proton. 
Accordingly, the enhancement
is seen as a signature for an $\eta^3$He bound state and a 
combined analysis including also $\gamma^3{\rm He} \rightarrow \eta^3{\rm He}$ 
data yielded a binding energy $B{=} 4.4{\pm}4.2$~MeV
and width $\Gamma{=}25.6{\pm}6.1$ MeV. Curiously enough
our prediction for $B$ presented above -- 
derived under the assumption that a bound state
exists -- is in line with this result and we would like to discuss it
a little further.
The solid squares in Fig.~\ref{bin4} show the difference between the
cross section of the reaction $\gamma^3{\rm He}{\to}\pi^0pX$ for
$170^0{\le}\theta_{\pi^0p}{\le}180^0$ and that for
$150^0{\le}\theta_{\pi^0p}{\le}170^0$
as a function of the invariant collision energy
$\sqrt{s}$. This difference exhibits an enhancement
in the vicinity of the $\gamma^3{\rm He}{\to}\eta^3{\rm He}$
threshold, the latter being indicated by the arrow.
Considering this enhancement to be entirely due to the formation
of a $\eta^3$He bound state we adopt the same strategy as in Ref. \cite{Pfeiffer}
and fit this resonant cross section, i.e. solid squares in 
Fig.~\ref{bin4}, using the nonrelativistic Breit-Wigner form
\begin{eqnarray}
\sigma =\frac{\Gamma^2/4}{(\sqrt{s}-m_\eta-m_{^3{\rm He}}+B)^2
+\Gamma^2/4},
\label{Breit}
\end{eqnarray}
with $m_\eta$=547.3 MeV, $m_{^3{\rm He}}$=2808.398 MeV.
The fit is shown by the dashed line in Fig.~\ref{bin4}.
We obtain $B$=0.43$\pm$2.9~MeV and $\Gamma$=27.9$\pm$7.2 MeV 
with a $\chi^2$/dof=0.89 based on the data points (squares)
shown in Fig.~\ref{bin4}.  Obviously our result is
smaller than the values quoted in Ref. \cite{Pfeiffer} 
and, as a matter of fact, so close to the threshold that it is 
compatible with zero. Thus, it is impossible to conclude from
these data alone whether the structure is a signature of a
bound state or not.
To improve the analysis one might combine the
$\gamma^3{\rm He}{\to}\pi^0pX$ and $\gamma^3{\rm He}{\to}\eta^3{\rm He}$
data as was done in Ref.~\cite{Pfeiffer}. The latter data are in
line with our analysis of the $\eta^3{\rm He}$ scattering length, 
cf. Fig.~\ref{bin3}, and accordingly with the limits we set for the
binding energy. However, as already mentioned above, these data do not
provide any constraints on the sign of $\Re a_{\eta{\rm He}}$. 
The sign is solely inferred from the data on $\gamma^3{\rm He}{\to}\pi^0pX$
and therefore subject to the ambiguity that is reflected in the large
uncertainty of the values for $B$ and $\Gamma$ that we (but also the authors of 
Ref.~\cite{Pfeiffer}) deduced. In this context we want to point out that
with the opening of the $\eta^3$He channel one would anyway expect
a cusp structure at the threshold which would give rise to a 
similar enhancement as exhibited by the dashed curve in Fig.~\ref{bin4},
but corresponds to a positive sign of $\Re a_{\eta{\rm He}}$.
Due to these
reasons we do not consider the results of Ref. \cite{Pfeiffer}
as an unambiguous signature of an $\eta^3$He bound state.

In summary, we have studied the relation between the $\eta^3$He
binding energy and width and the $\eta^3$He scattering length.
While, based on final state analyses of $\eta$ production
reactions, one cannot obtain direct information about the existence of 
$\eta^3$He bound states, it is, nevertheless, possible to find constraints
regarding the range of energies and widths where such a bound state
could be possible. Thus,
assuming that the real part of the $\eta^3$He scattering length
is negative and following our systematic
analysis~\cite{Sibirtsev} of  $a_{\eta{\rm He}}$
evaluated from available data for the $pd{\to}\eta^3$He reaction,
we set the limits for the binding energy to $B{\le}5$~MeV and
for the width to $\Gamma{\le}10$~MeV. 
However, whether or not a bound state indeed exists cannot be deduced
from our analysis, simply because we have no reliable empirical 
information on the sign of $\Re a_{\eta{\rm He}}$. 
 
\vskip -1cm 
\begin{acknowledgments}
We thank M. Pfeiffer for discussions of the TAPS data and C. Hanhart
for useful comments.
This work was supported by the DAAD and Academy of Finland
exchange programme projects 313-SF-PPP-8 (Germany) and 41926 (Finland) and
Academy of Finland grant number  54038. J.A.N. also thanks the Magnus
Ehrnrooth Foundation for partial support in this work.
\end{acknowledgments}


\begin{references}
\bibitem{Watson}
         K.M. Watson, Phys. Rev. {\bf 88}, 1163 (1952).
\bibitem{Migdal}
         A.B. Migdal, JETP {\bf 1}, 2 (1955).
\bibitem{Mayer}
         B. Mayer {\it et al.}, Phys. Rev. C {\bf 53}, 2068 (1996).
\bibitem{Green}
         A.M. Green and S. Wycech, Phys. Rev. C {\bf 68}, 061601 (2003).
\bibitem{Sibirtsev}
         A. Sibirtsev, J. Haidenbauer, C. Hanhart, and J. A. Niskanen,
         arXiv:nucl-th/0310079,  Eur. Phys. J. {\bf A},
         to be published.
\bibitem{Baru}
        V. Baru, J. Haidenbauer, C. Hanhart, and J. A. Niskanen,
        Phys. Rev. C {\bf 68}, 035203 (2003).
\bibitem{Wilkin}
        C. Wilkin, Phys. Rev. C. {\bf 47}, R938 (1993).
\bibitem{Wycech}
        S. Wycech, A.M. Green and J.A. Niskanen, Phys. Rev.
        C {\bf 52}, 544 (1995).
\bibitem{Rakityansky}
        S.A. Rakityansky,S.A. Sofianos, M. Braun, V.B. Belyaev
        and W. Sandhas, Phys. Rev. C {\bf 53}, R2043 (1996).
\bibitem{Fix1}
        A. Fix and H. Arenh\"ovel, Phys. Rev. C {\bf 66}, 024002 (2002).
\bibitem{Haider1}
        Q. Haider and L.C. Liu, Phys. Lett. B {\bf 172}, 257 (1986).
\bibitem{Liu}
        L.C. Liu and Q. Haider, Phys. Rev. C {\bf 34}, 1845 (1986).
\bibitem{Chiang}
        H.C. Chiang, E. Oset and L.C. Liu, Phys. Rev. C {\bf 44}, 738 (1991).
\bibitem{Haider2}
        Q. Haider and L.C. Liu, Phys. Rev. C {\bf 66}, 045208 (2002).
\bibitem{Scoccola}
        N. N. Scoccola and D.-O. Riska, Phys. Lett. B {\bf 444}, 21 (1998).
\bibitem{cracow}
  A. Sibirtsev, J. Haidenbauer, C. Hanhart, and J. A. Niskanen,
  Proc. Int. Conf. on Meson-Nucleon Physics MENU2004, Cracow, Poland,
  June 2004, (World Scientific, Singapore), to be published.
\bibitem{Joachain} C. J. Joachain, {\it Quantum Collision Theory},
(North Holland, Amsterdam, 1975).

\bibitem{Berger}
        J. Berger et al., Phys. Rev. Lett. {\bf 61}, 919 (1988).
\bibitem{Pfeiffer}
        M. Pfeiffer et al., Phys. Rev. Lett. {\bf 92},
        252001 (2004).
\bibitem{Banaigs}
        J. Banaigs et al., Phys. Lett. B {\bf 45}, 394 (1973).
\bibitem{Bilger}
        R. Bilger et al., Phys. Rev. C {\bf 65}, 044608 (2002).
\bibitem{Betigeri}
        M. Betigeri et al, Phys. Lett. B {\bf 472}, 267 (2000).
\bibitem{Kirchner}
        T. Kirchner, PhD. Thesis, Institute de Physique Nucleaire,
        Orsay (1993).
\end{references}
\end{document}